\documentclass[iop,apj]{emulateapj}
\usepackage{epsf}

\newcommand{\etal}{{et al.~}}
\newcommand{\masyr}{ \ {\rm{mas \ yr^{-1}}}\>}
\newcommand{\kms}{ \ {\rm{km \ s^{-1}}}\>}

\newcommand{\kpc}{\>{\rm kpc}}
\newcommand{\Myr}{\>{\rm Myr}}

\begin{document}

\pagenumbering{arabic}

\title{First Gaia Local Group Dynamics:
Magellanic Clouds Proper Motion and Rotation}

\author{Roeland P.~van der Marel}
\affil{Space Telescope Science Institute, 3700 San Martin Drive, 
       Baltimore, MD 21218}
\author{Johannes Sahlmann}
\affil{European Space Agency, Space Telescope Science Institute, 
       3700 San Martin Drive, Baltimore, MD 21218, USA}


\slugcomment{ApJ Letters, submitted, September 14, 2016}
\shorttitle{Gaia Magellanic Clouds Proper Motion and Rotation}
\shortauthors{van der Marel \& Sahlmann}

\begin{abstract}
We use the \emph{Gaia} data release 1 (DR1) to study the proper motion
(PM) fields of the Large and Small Magellanic Clouds (LMC, SMC). This
uses the \emph{Tycho}-\emph{Gaia} Astrometric Solution (TGAS) PMs for
29 \emph{Hipparcos} stars in the LMC and 8 in the SMC. The LMC PM in
the West and North directions is inferred to be $(\mu_W,\mu_N) =
(-1.872 \pm 0.045, 0.224 \pm 0.054) \masyr$, and the SMC PM
$(\mu_W,\mu_N) = (-0.874 \pm 0.066, -1.229 \pm 0.047) \masyr$. These
results have similar accuracy and agree to within the uncertainties
with existing Hubble Space Telescope (\emph{HST}) PM measurements.
Since TGAS uses different methods with different systematics, this
provides an external validation of both data sets and their underlying
approaches. Residual DR1 systematics may affect the TGAS results, but
the \emph{HST} agreement implies this must be below the random
errors. Also in agreement with prior \emph{HST} studies, the TGAS LMC
PM field clearly shows the clockwise rotation of the disk, even though
it takes the LMC disk in excess of $10^8$ years to complete one
revolution. The implied rotation curve amplitude for young LMC stars
is consistent with that inferred from line-of-sight (LOS) velocity
measurements. Comparison of the PM and LOS rotation curves implies a
kinematic LMC distance modulus $m-M = 18.54 \pm 0.39$, consistent but
not yet competitive with photometric methods. These first results from
\emph{Gaia} on the topic of Local Group dynamics provide an indication
of how its future data releases will revolutionize this field.
\end{abstract}


\section{Introduction}
\label{s:intro}

Almost everything that is known about Local Group dynamics, and of
galaxy dynamics in general, is based on LOS velocity
observations. Such measurements constrain only one component of
motion, and interpretation therefore requires that various assumptions
be made.  PMs in the plane of the sky provide a more complete
picture. However, the PMs are generally small and inversely
proportional to the distance of the target.

The \emph{Hipparcos} satellite provided a detailed understanding of
the PMs of stars in the solar neighborhood (ESA 1997), but its
accuracy was insufficient for detailed studies of other Local Group
objects. Water maser observations yielded the first accurate PMs for
other Local Group galaxies (Brunthaler et al.~2005). However, this
technique is limited to a few galaxies with high star formation rates.
Only with \emph{HST} has it become possible to determine PMs for
objects throughout the Local Group. For example, the HSTPROMO
collaboration has studied the PM dynamics of globular clusters,
stellar streams, and nearby galaxies (van der Marel 2015).

The \emph{Gaia} satellite will provide the next step forward through
PM measurements for objects across the sky to optical magnitude $\sim$20.7
(Gaia Collaboration 2016a). Initial five-parameter astrometric
solutions (including PMs) are expected in late 2017, with a final DR
in 2022. By contrast, the \emph{Gaia} DR1 of Sep 14, 2016 (Gaia
Collaboration \etal 2016b), includes PMs only for stars in common
between \emph{Gaia} and the \emph{Hipparcos} Tycho-2 Catalogue (Hoeg
\etal 2000). This Tycho-\emph{Gaia} Astrometric Solution (TGAS)
Catalog (Lindegren \etal 2016) is restricted to the same bright stars
previously studied by \emph{Hipparcos} and is therefore not well
suited for studies of Local Group dynamics. However, we show in this paper that
it does yield some first new insights.

The LMC and SMC are the most massive satellites of the Milky Way. They
have been studied extensively for a wide range of astrophysical
subjects. To place these results in a proper context, it is important
to understand the dynamics of the Magellanic Clouds and their history
in the Local Group. They have therefore been of special interest for
\emph{HST} studies. Kallivayalil \etal (2006a,b) presented PMs for 26
fields based on two epochs of \emph{HST} data with a 2-year time
baseline. These measurements were refined by Kallivayalil \etal (2013;
hereafter K13) using a third epoch for 12 fields, which extended the
time baseline to 7 years. The latter provided a median per-coordinate
PM uncertainty of only 0.03 mas/yr (7 km/s), 3--4 times better than
the two-epoch measurements.


\begin{deluxetable*}{rrrrrr}
\setlength{\tablewidth}{\hsize}
\tablecaption{TGAS Proper Motions of Magellanic Cloud stars\label{t:TGAS}}
\tablehead{
\colhead{HIP} & \colhead{\emph{Gaia}} & \colhead{RA} & \colhead{dec} & 
\colhead{${\rm PM}_W$} & \colhead{${\rm PM}_N$} \\
\colhead{ID}  & \colhead{sourceId}   & \colhead{(deg)} & \colhead{(deg)} & \colhead{(mas/yr)} &
\colhead{(mas/yr)} \\
} 
\startdata
LMC\ \ \\
\hline
22392 & 4655349652394811136 & 72.3017 & -69.4565 & $-2.012 \pm 0.151$ & $-0.226 \pm 0.151$ \\
22758 & 4655510043652327552 & 73.4304 & -68.7148 & $-1.772 \pm 0.160$ & $-0.296 \pm 0.171$ \\
22794 & 4655460771785226880 & 73.5594 & -69.2101 & $-1.895 \pm 0.088$ & $-0.122 \pm 0.088$ \\
22849 & 4661769941306044416 & 73.7390 & -66.7524 & $-1.756 \pm 0.120$ & $-0.045 \pm 0.130$ \\
22885 & 4661720532007512320 & 73.8400 & -67.4365 & $-1.766 \pm 0.145$ & $-0.030 \pm 0.147$ \\
22900 & 4655136518933846784 & 73.8853 & -69.9625 & $-1.951 \pm 0.232$ & $-0.102 \pm 0.221$ \\
22989 & 4655158131209278464 & 74.1962 & -69.8402 & $-1.869 \pm 0.234$ & $0.015 \pm 0.227$ \\
23177 & 4662293892954562048 & 74.7878 & -65.6677 & $-1.613 \pm 0.162$ & $0.026 \pm 0.159$ \\
23428 & 4654621500815442816 & 75.5308 & -71.3370 & $-1.973 \pm 0.146$ & $-0.099 \pm 0.140$ \\
23527 & 4655036841335115392 & 75.8733 & -70.6998 & $-1.942 \pm 0.240$ & $0.077 \pm 0.230$ \\
23665 & 4661920986713556352 & 76.3009 & -66.7368 & $-1.675 \pm 0.104$ & $0.005 \pm 0.114$ \\
23718 & 4661472145451256576 & 76.4813 & -67.8864 & $-1.785 \pm 0.072$ & $0.123 \pm 0.079$ \\
23820 & 4662061311885050624 & 76.8091 & -66.0551 & $-1.555 \pm 0.212$ & $0.238 \pm 0.216$ \\
24006 & 4651629489160555392 & 77.4122 & -71.4006 & $-2.236 \pm 0.105$ & $0.065 \pm 0.098$ \\
24080 & 4658269336800428672 & 77.5950 & -68.7733 & $-1.896 \pm 0.092$ & $0.169 \pm 0.094$ \\
24347 & 4658204053297963392 & 78.3783 & -69.5399 & $-2.084 \pm 0.196$ & $0.251 \pm 0.177$ \\
24694 & 4658137739001073280 & 79.4433 & -69.8492 & $-1.882 \pm 0.131$ & $0.182 \pm 0.126$ \\
24988 & 4660601607121368704 & 80.2571 & -65.8007 & $-1.499 \pm 0.079$ & $0.387 \pm 0.089$ \\
25097 & 4660444926713007872 & 80.5878 & -66.2603 & $-1.510 \pm 0.173$ & $0.337 \pm 0.191$ \\
25448 & 4658486455992620416 & 81.6454 & -68.8687 & $-1.710 \pm 0.138$ & $0.587 \pm 0.133$ \\
25615 & 4660175580731856128 & 82.0847 & -67.4051 & $-1.568 \pm 0.204$ & $0.479 \pm 0.208$ \\
25892 & 4660124762671796096 & 82.9101 & -67.4699 & $-1.587 \pm 0.182$ & $0.669 \pm 0.186$ \\
26135 & 4660246224352015232 & 83.5936 & -67.0232 & $-1.633 \pm 0.095$ & $0.429 \pm 0.113$ \\
26222 & 4657280635327480832 & 83.8193 & -69.6773 & $-1.723 \pm 0.187$ & $0.497 \pm 0.188$ \\
26338 & 4657700408260606592 & 84.1349 & -68.9005 & $-1.874 \pm 0.185$ & $0.621 \pm 0.200$ \\
26745 & 4657627943562907520 & 85.2409 & -69.2586 & $-1.779 \pm 0.231$ & $0.518 \pm 0.242$ \\
27142 & 4657722879521554176 & 86.3193 & -68.9978 & $-1.733 \pm 0.142$ & $0.705 \pm 0.137$ \\
27819 & 4659188769038018816 & 88.2918 & -68.1186 & $-1.560 \pm 0.153$ & $0.834 \pm 0.106$ \\
27868 & 4659091084305723392 & 88.4571 & -68.3132 & $-1.661 \pm 0.154$ & $0.843 \pm 0.111$ \\
\hline
\multicolumn{2}{l}{straight mean\tablenotemark{[1]}} & & & $-1.776 \pm 0.033$ & $ 0.246 \pm 0.059$ \\
\multicolumn{2}{l}{weighted mean\tablenotemark{[2]}} & & & $-1.779 \pm 0.024$ & $ 0.241 \pm 0.024$ \\
\hline\\
SMC\ \ \\
\hline
3934 & 4685876046561549184 & 12.6316 & -73.4785 & $-0.541 \pm 0.177$ & $-1.304 \pm 0.177$ \\
3945 & 4685876046561548800 & 12.6600 & -73.4717 & $-0.668 \pm 0.154$ & $-1.160 \pm 0.148$ \\
4004 & 4689033534707612800 & 12.8525 & -72.3829 & $-0.670 \pm 0.148$ & $-1.165 \pm 0.143$ \\
4126 & 4685940436697751168 & 13.2135 & -73.1149 & $-0.667 \pm 0.132$ & $-1.291 \pm 0.116$ \\
4153 & 4688967357860689024 & 13.2704 & -72.6334 & $-0.821 \pm 0.131$ & $-1.231 \pm 0.130$ \\
4768 & 4690499767820637312 & 15.3208 & -72.2920 & $-1.144 \pm 0.151$ & $-1.239 \pm 0.143$ \\
5267 & 4687436700227349888 & 16.8259 & -72.4677 & $-0.849 \pm 0.152$ & $-1.262 \pm 0.144$ \\
5714 & 4687159863816994816 & 18.3771 & -73.3362 & $-0.992 \pm 0.091$ & $-1.182 \pm 0.082$ \\
\hline
\multicolumn{2}{l}{straight mean\tablenotemark{[1]}} & & & $-0.794 \pm 0.066$ & $ -1.229 \pm 0.018$ \\
\multicolumn{2}{l}{weighted mean\tablenotemark{[2]}} & & & $-0.830 \pm 0.047$ & $ -1.222 \pm 0.044$
\enddata
\tablecomments{Column~(1)-(2): \emph{Hipparcos} ID and \emph{GAIA}
  sourceId number of those Magellanic Cloud stars previously
  identified by Kroupa \& Bastian (1997) and analyzed
  here. Columns~(3)-(6): right ascension $\alpha$, declination
  $\delta$, proper motion ${\rm PM}_{\rm W(est)}$ ($\equiv - {\rm
    PM}_\alpha \cos \delta$), and ${\rm PM}_{\rm N(orth)}$ ($\equiv
  {\rm PM}_\delta$) from the TGAS catalog. All stars have known LOS
  velocities (Barbrier-Brossat \etal 1994; Kroupa \& Bastian 1997;
  Neugent \etal 2012; Kordopatis \etal 2013) consistent with LMC or
  SMC membership. The stars HIP 22237 and 25815 in the LMC were
  rejected because they had the two highest TGAS {\tt
    astrometric\_excess\_noise} values ($> 1.02$) in the sample, as
  well as strongly outlying PM values. HIP 7912 and 8470 near the SMC
  were excluded because they reside in the Magellanic Bridge. HIP
  23500, 24907, 25146, 25822, and 27655 in the LMC, and HIP 5397 in
  the SMC are not listed in the TGAS catalog. There are no additional
  \emph{Hipparcos} stars with both position and kinematics (SIMBAD LOS
  velocity and TGAS PM) consistent with Magellanic Cloud
  membership. The straight and weighted mean for each Magellanic Cloud
  are listed (accounting also for correlations between the TGAS PM
  components).}

\tablenotetext{[1]}{The uncertainty in the ``straight'' mean is based
  exclusively on the observed scatter, and doesn't use the individual
  PM uncertainties.}

\tablenotetext{[2]}{The uncertainty in the weighted mean is based
  exclusively on the individual PM uncertainties, and doesn't use the
  observed scatter. This underestimates the PM uncertainty in the COM
  motion of each Magellanic Cloud.}
\end{deluxetable*}

The \emph{HST} studies showed that the Magellanic Clouds move faster
about the Milky Way than previously believed based on models of the
Magellanic Stream. So instead of being long-term satellites, they are
most likely on their first Milky Way passage (Besla \etal 2007). These
results have refined our understanding of the Magellanic Clouds, as
well as the formation of Magellanic Irregulars in general (Besla \etal
2012). van der Marel \& Kallivayalil (2014; hereafter vdMK14) studied
the variations in the \emph{HST} PM measurements across the face of
the LMC. They measured the PM rotation curve, and demonstrated
consistency with LOS velocity studies.


\begin{figure*}[t]
\begin{center}
\epsfxsize=0.49\hsize
\epsfbox{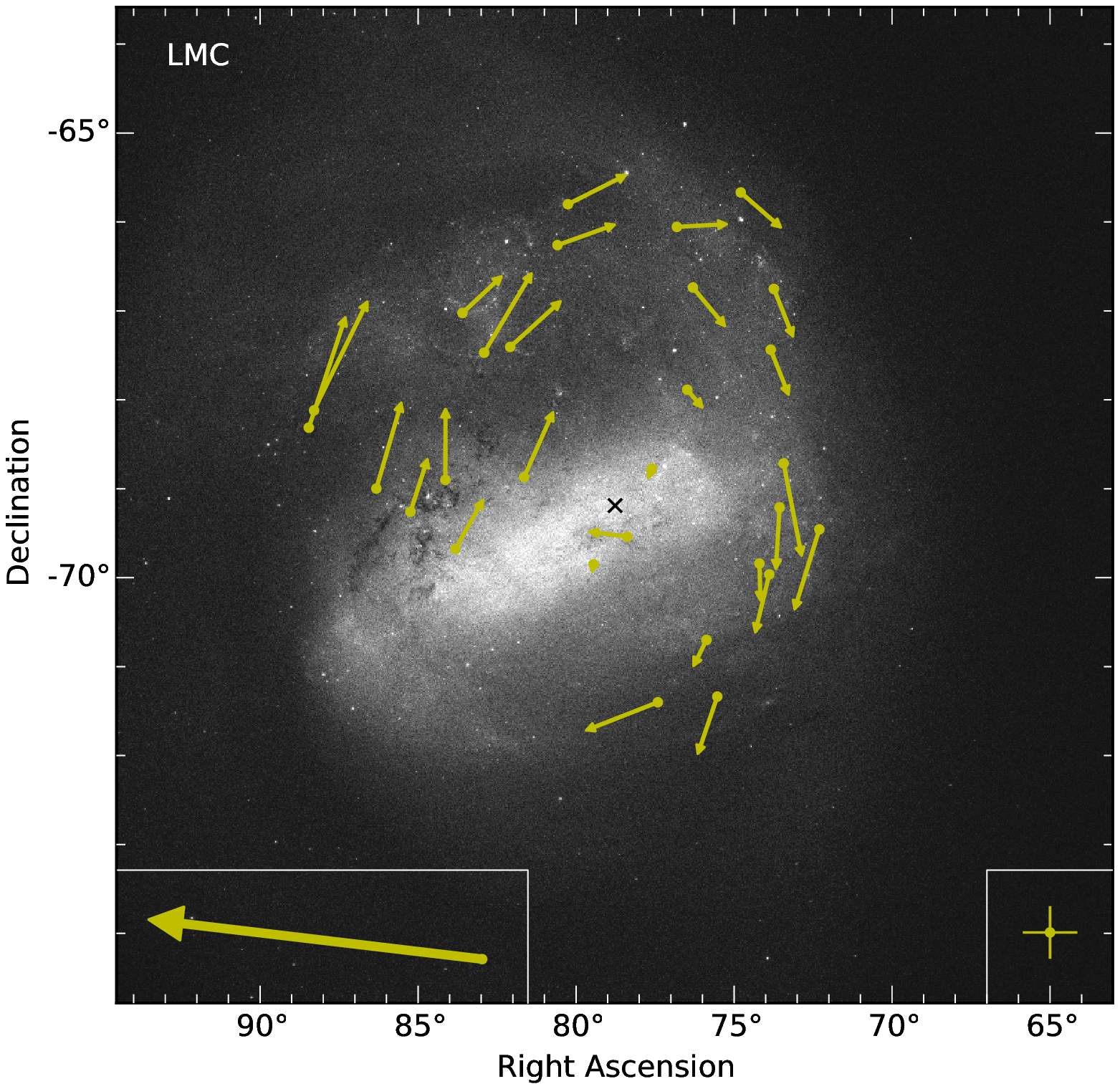}
\hfill
\epsfxsize=0.49\hsize
\epsfbox{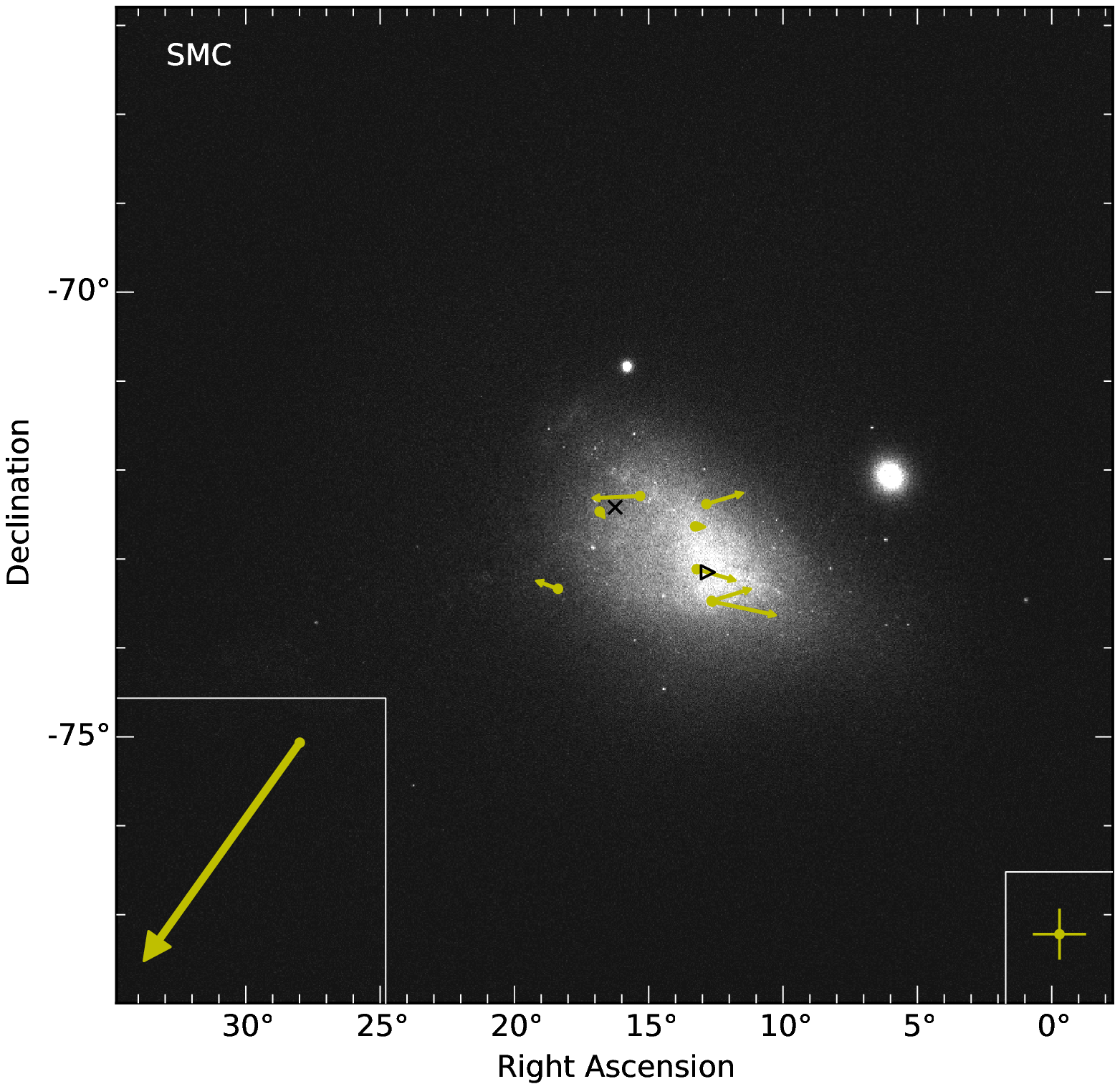}
\caption{Spatially variable component of the observed TGAS PM fields
  for the LMC (left) and SMC (right), overlaid on a representation of
  the Gaia DR1 source density. Each panel is centered on the dynamical
  center (cross; a triangle for the SMC indicates the photometric
  center of the old stars), with the horizontal and vertical extent
  representing an equal number of degrees on the sky.  Solid dots show
  the positions of the sample stars. The PM vector for each star is
  the observed PM from Table~\ref{t:TGAS}, minus the best-fit COM PM
  from Tables~\ref{t:param} and~\ref{t:paramS} (bottom left inset). PM
  vectors have a size that indicates the mean predicted motion over
  (arbitrarily) the next $7.2 \Myr$. For the LMC, clockwise rotation
  is clearly evident. The bottom right inset shows the median random
  PM errors for the sample.\footnote{A figure that shows the
    individual PM uncertainties is available at
    http://www.cosmos.esa.int/web/gaia/iow\_20160916 .} The $\chi^2$
  values of our model fits suggest that these may be overestimated by
  a factor $\sim 1.85$ (confirmed visually by the good agreement
  between the PMs of adjacent stars). Overstimated errors have also
  been suggested for Gaia DR1 parallax values (Casertano \etal 2016).
\label{f:obsvar}}
\end{center}
\end{figure*}

Historically, one of the first measurements of the Magellanic Cloud
PMs was obtained by Kroupa \& Bastian (1997), using data for 36 LMC
stars and 11 SMC stars from the \emph{Hipparcos} satellite. These are
young massive stars with apparent V-magnitudes between 9--12 (absolute
magnitudes brighter than $-6.5$). High-quality TGAS data exist for 29
of the LMC and 8 of the SMC stars. We retrieved these data from the
\emph{Gaia} archive using
\texttt{pygacs}.\footnote{\url{https://github.com/Johannes-Sahlmann/pygacs}}
While the \emph{Hipparcos} PM errors ranged from one to a few mas/yr,
the new TGAS PM errors, listed in Table~\ref{t:TGAS}, are much
smaller. The 0.15 mas/yr median error is similar to the \emph{HST} PM
errors for the K13 two-epoch fields. So while the TGAS measurements do
not improve upon the \emph{HST} measurements, they do allow for an
independent verification.  We therefore analyze here the Magellanic
Cloud TGAS PMs with the same methodologies presented in K13 and
vdMK14. We do this for the LMC in Section~2, and the SMC in
Section~3. Section~4 discusses the results in the context of previous
work. Section 5~summarizes the conclusions.

\section{LMC Proper Motion Field}
\label{s:pmanalysis}

Figure~\ref{f:obsvar}a shows the spatially variable component of the
observed LMC PM field (comparable to Figure~1 of vdMK14 for the
\emph{HST} PM data). For each star in Table~\ref{t:TGAS} we subtracted
the best-fit LMC center-of-mass (COM) PM (left inset) derived
below. Clockwise motion is clearly evident. This qualitatively
validates the accuracy of the data, and confirms that the stars belong
to the LMC.

We model the LMC PM field to derive its kinematical and geometrical
parameters. The model has contributions from the internal rotation of
the LMC and the systemic motion of the LMC COM. To describe the
former, we assume that the LMC is a flat disk with circular
streamlines. The latter adds a spatially variable ``perspective
rotation'' component (due to the fact that projection of the COM
velocity vector onto the West and North directions depends on
position). The relation between the transverse velocity $v_t$ in km/s
and the PM $\mu$ in mas/yr is given by $\mu = v_t / (4.74047 D)$,
where $D$ is the distance in kpc (since the LMC is an inclined disk,
this distance $D$ is not the same for all stars). We refer the reader
to van der Marel \etal (2002; hereafter vdM02) for a derivation of
full analytical expressions.

Table~\ref{t:param} lists the results of fitting this model to the
data, accounting for the small correlations between the TGAS PM
components.\footnote{The average correlation between ${\rm PM}_W$ and
  ${\rm PM}_N$ is smaller than 0.1 for the LMC TGAS stars.} Column~(2)
uses only the new TGAS PM data. The COM position is not
well-constrained by the data, since there are very few TGAS stars on
the south-east side of the LMC; we therefore keep it fixed at the
value inferred by vdMK14. Column~(3) is the fit to the \emph{HST} PM
data from vdMK14, which pertain to a mix of young and old stellar
populations with apparent V-magnitudes between 16--24. Column~(4) fits
both data sets together. This improves the constraints, but
complicates the interpretation by mixing stars of different ages
(which have different kinematics because of the phenomenon of
asymmetric drift). Column~(5) shows results obtained when the TGAS PMs
are fit simultaneously with an age-matched sample of literature LOS
velocities for 723 Red Supergiants (vdMK14).


\begin{deluxetable*}{llrrr}
\setlength{\tabcolsep}{20pt}
\setlength{\tablewidth}{\hsize}
\tablecaption{LMC Dynamical Model Parameters: Fit Results from Two- and 
Three-Dimensional Kinematics\label{t:param}}
\tablehead{
\colhead{Quantity (unit)} & \colhead{PMs} & \colhead{PMs} & \colhead{PMs} & \colhead{PMs TGAS +}\\
\colhead{} & \colhead{TGAS} & \colhead{HST} & \colhead{TGAS+HST}
 & \colhead{Young Star $v_{\rm LOS}$}\\
\colhead{(1)} & \colhead{(2)} & \colhead{(3)} & \colhead{(4)} & \colhead{(5)}}
\startdata
$\alpha_0$ (deg)              & $78.76 \pm 0.52$\tablenotemark{[1]} 
                              & $78.76 \pm 0.52$
                              & $79.37 \pm 0.75$       
                              & $80.21 \pm 0.40$\\
$\delta_0$ (deg)              & $-69.19 \pm  0.25$\tablenotemark{[1]}
                              & $-69.19 \pm  0.25$
                              & $-69.58 \pm 0.38$
                              & $-69.26 \pm 0.20$\\
$i$ (deg)                     & $ 43.7 \pm 8.3$     
                              & $ 39.6 \pm 4.5$
                              & $ 37.7 \pm 5.2$
                              & $ 30.3 \pm 5.9$\\
$\Theta$ (deg)                & $124.2 \pm 28.7$                   
                              & $147.4 \pm 10.0$
                              & $142.3 \pm 9.7$
                              & $153.7 \pm 5.4$\\
$\mu_{W0}$ (mas/yr)           & $-1.872 \pm 0.045$   
                              & $-1.910 \pm 0.020$
                              & $-1.905 \pm 0.023$
                              & $-1.850 \pm 0.031$\\
$\mu_{N0}$ (mas/yr)           & $0.224 \pm 0.054$   
                              & $0.229 \pm 0.047$
                              & $0.275 \pm 0.050$
                              & $0.350 \pm 0.035$\\
$v_{\rm LOS,0}$ (km/s)        & $262.2 \pm 3.4$\tablenotemark{[1]}
                              & $262.2 \pm 3.4$\tablenotemark{[1]}
                              & $262.2 \pm 3.4$\tablenotemark{[1]}
                              & $270.5 \pm 3.0$\\
$V_0$ (km/s)                  & $107.1 \pm 22.4$
                              & $ 76.1 \pm  7.6$
                              & $ 78.9 \pm  7.5$
                              & $ 77.5 \pm 14.7$\\
$R_0/D_0$                     & $0.087 \pm 0.019$
                              & $0.024 \pm 0.010$
                              & $0.052 \pm 0.010$
                              & $0.045 \pm 0.005$\\
$D_0$ (kpc)                   & $50.1 \pm  2.5 \kpc$\tablenotemark{[1]}
                              & $50.1 \pm  2.5 \kpc$\tablenotemark{[1]}
                              & $50.1 \pm  2.5 \kpc$\tablenotemark{[1]}
                              & $51.1 \pm  9.5 \kpc$
\enddata

\tablecomments{Columns~(1): model quantity and units.
  Columns~(2)-(5): values inferred from model fits described in
  Section~\ref{s:pmanalysis}. The listed quantities are: position
  $(\alpha_0,\delta_0)$ of the dynamical center; inclination angle $i$
  and line-of-nodes position angle $\Theta$ of the disk plane; COM PM
  $(\mu_{W0},\mu_{N0})$ and LOS velocity $v_{\rm LOS,0}$; rotation
  curve amplitude $V_0$ and turnover radius $R_0$; and distance
  $D_0$.}

\tablenotetext{[1]}{$(\alpha_0,\delta_0)$ from vdMK14; $v_{\rm LOS,0}$
  from vdM02; $D_0$ based on distance modulus $m-M = 18.50 \pm 0.10$
  from Freedman \etal (2001). Not fit, but uncertainty propagated into
  other model parameters.}
\end{deluxetable*}

\begin{figure}
\begin{center}
\epsfxsize=\hsize
\centerline{\epsfbox{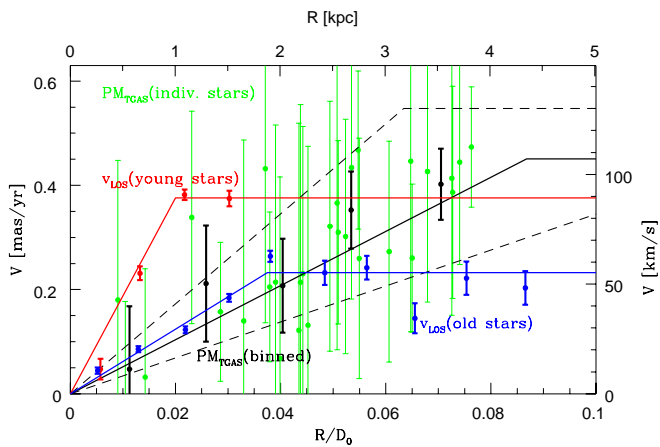}}
\caption{LMC disk rotation velocity $V$ at cylindrical radius $R$.
  The left and bottom axes are expressed in angular and dimensionless
  units, while the right and top axes show physical units. Errorbars
  include only the random measurement noise, not propagated
  uncertainties from other LMC model parameters. Broken curves shows
  the best-fit parameterizations of the form used in
  Section~\ref{s:pmanalysis}. Red and blue data points show LOS
  rotation curves for samples of young and old stars (vdMK14). Green
  data points show the PM rotation curve inferred here from individual
  TGAS stars. Black data points show a radial binning of the TGAS
  results in $0.8 \kpc$ bins, yielding 3, 2, 9, 8, and 7 stars in the
  subsequent bins, respectively. The binned data for $V(R)$, with $V$
  in km/s and $R$ in kpc, are: $V(0.56) = 11.2 \pm 28.7$, $V(1.29) =
  50.3 \pm 26.5$, $V(2.03) = 49.2 \pm 21.4$, $V(2.68) = 83.8 \pm 17.6$,
  and $V(3.53) = 95.5 \pm 16.2$.\label{f:rotcurve}}
\end{center}
\end{figure}

These fits parameterize the rotation curve $V(R)$ as function of radius
$R$ in the disk to increase linearly to velocity $V_0$ at radius
$R_0$, and then stay flat at larger radii. We also obtained a
non-parametric estimate for $V(R)$ as in vdMK14, by determining for
each star the PM component along the local direction of
rotation. Green data points in Figure~\ref{f:rotcurve} show the
rotation estimates thus obtained for the TGAS stars, while black data
points show the results after binning in radius to decrease the error
bars.


\section{SMC Proper Motion Field}
\label{s:pmanalysisS}

Figure~\ref{f:obsvar}b shows the spatially variable component of the
observed SMC PM field, after subtraction of the best-fit COM PM
derived below. The stars all have similar PMs, which confirms their
SMC membership. No rotation in the plane of the sky is evident. This
is due to two separate effects.  First, the SMC is smaller than the
LMC, and the TGAS stars are closer to the galaxy center then they are
for the LMC. At small radii, both the intrinsic galaxy rotation and
perspective rotation components are smaller. Second, photometric and
LOS velocity studies show that the SMC is more vertically extended and
less rapidly rotating (if at all) than the LMC (van der Marel \etal
2009).

In view of these facts and the small number of stars, we fit a
relatively simple model to the SMC PM field, as in K13. We include
perspective rotation and allow for a single overall intrinsic galaxy
rotation velocity $V_{\rm rot}$ in the plane of the sky (i.e., as
though we were viewing a face-on disk). We keep the distance modulus
fixed at $m-M = 18.99 \pm 0.10$ (Cioni \etal 2000b), and the radial
velocity fixed at $v_{\rm sys} = 145.6 \pm 0.6$ km/s (Harris \&
Zaritsky 2006).

We explored several different fits, in which the SMC COM PM is always
a free parameter. We keep the COM position fixed, since it is not well
constrained by the PM data in absence of rotation. We use either the
HI kinematical center from Stanimirovic \etal (2014), or the
photometric center of the old stars from Cioni \etal (2000a), in each
case with an uncertainty of $0.2^{\circ}$ per coordinate. We treat
$V_{\rm rot}$ either as a free parameter, or keep it fixed to $V_{\rm
  rot} = 0 \pm 40 \kms$. The assigned uncertainty is the rotation
velocity of HI in the SMC (Stanimirovic \etal 2014), which should
exceed the amount of rotation in the young stellar
component. Table~\ref{t:paramS} lists the results of fitting the TGAS
PM data by themselves, the \emph{HST} PM data (K13) by themselves, or
both data sets together.


\begin{deluxetable*}{llrrrr}
\setlength{\tabcolsep}{20pt}
\setlength{\tablewidth}{\hsize}
\tablecaption{SMC Dynamical Model Parameters: Fit Results from 
Two-Dimensional Kinematics\label{t:paramS}}
\tablehead{
\colhead{Quantity} & \colhead{Unit} & \colhead{$V_{\rm rot}$ varied} & 
\colhead{$V_{\rm rot}$ varied} & \colhead{$V_{\rm rot}$ fixed} & \colhead{$V_{\rm rot}$ fixed} \\
\colhead{} & \colhead{} & \colhead{center: HI} & \colhead{center: old stars} &
\colhead{center: HI} & \colhead{center: old stars} \\
\colhead{(1)} & \colhead{(2)} & \colhead{(3)} & \colhead{(4)} 
& \colhead{(5)} & \colhead{(6)} } 
\startdata
$\alpha_0$          &  deg    & $16.25 \pm 0.20$\tablenotemark{[1]} 
                              & $12.80 \pm 0.20$\tablenotemark{[1]}
                              & $16.25 \pm 0.20$\tablenotemark{[1]}
                              & $12.80 \pm 0.20$\tablenotemark{[1]}\\
$\delta_0$          &  deg    & $-72.42 \pm 0.20$\tablenotemark{[1]}
                              & $-73.15 \pm 0.20$\tablenotemark{[1]}
                              & $-72.42 \pm 0.20$\tablenotemark{[1]}
                              & $-73.15 \pm 0.20$\tablenotemark{[1]}\\
\hline
PMs TGAS \\
\hline
$\mu_{W0}$          &  mas/yr & $-0.874 \pm 0.066$   
                              & $-0.777 \pm 0.058$
                              & $-0.852 \pm 0.104$
                              & $-0.790 \pm 0.071$\\
$\mu_{N0}$          &  mas/yr & $-1.229 \pm 0.047$   
                              & $-1.211 \pm 0.070$   
                              & $-1.215 \pm 0.091$   
                              & $-1.256 \pm 0.107$\\
$V_0$               &  km/s   & $-15.2 \pm 16.8$
                              & $-22.7 \pm 19.2$
                              & $ 0.0 \pm 40.0$\tablenotemark{[1]}
                              & $ 0.0 \pm 40.0$\tablenotemark{[1]}\\
\hline
PMs HST \\
\hline
$\mu_{W0}$          &  mas/yr & $-0.694 \pm 0.082$   
                              & $-0.722 \pm 0.041$
                              & $-0.772 \pm 0.125$
                              & $-0.715 \pm 0.077$\\
$\mu_{N0}$          &  mas/yr & $-1.055 \pm 0.068$   
                              & $-1.171 \pm 0.074$   
                              & $-1.117 \pm 0.109$   
                              & $-1.153 \pm 0.101$\\
$V_0$               &  km/s   & $ 29.5 \pm 25.0$
                              & $  7.7 \pm 22.9$
                              & $ 0.0 \pm 40.0$\tablenotemark{[1]}
                              & $ 0.0 \pm 40.0$\tablenotemark{[1]}\\
\hline
PMs TGAS+HST \\
\hline
$\mu_{W0}$          &  mas/yr & $-0.819 \pm 0.041$   
                              & $-0.733 \pm 0.031$
                              & $-0.799 \pm 0.102$
                              & $-0.740 \pm 0.072$\\
$\mu_{N0}$          &  mas/yr & $-1.177 \pm 0.039$   
                              & $-1.185 \pm 0.042$   
                              & $-1.164 \pm 0.081$   
                              & $-1.202 \pm 0.107$\\
$V_0$               &  km/s   & $ -8.8 \pm 12.3$
                              & $ -8.1 \pm 15.2$
                              & $ 0.0 \pm 40.0$\tablenotemark{[1]}
                              & $ 0.0 \pm 40.0$\tablenotemark{[1]}
\enddata

\tablecomments{Column~(1)-(2): model quantity and units, defined
  similarly as in Table~\ref{t:param}. Columns~(3)-(6): values
  inferred from model fits described in Section~\ref{s:pmanalysisS}.}

\tablenotetext{[1]}{Value kept fixed. Not fit, but uncertainty
  propagated into other model parameters.}
\end{deluxetable*}

\section{Discussion}
\label{s:disc}

The results obtained here are generally consistent with what has been
previously reported in the literature, as summarized in K13 and
vdMK14.
 
The TGAS value for the LMC COM PM (Table~\ref{t:param}, column~(2))
can be compared to the \emph{HST} measurement in
column~(3). Similarly, the TGAS value for the SMC COM PM can be
compared to the corresponding \emph{HST} measurement for any given set
of model assumptions Table~\ref{t:paramS}. The results from TGAS and
\emph{HST} have similar random errors, and the COM PM measurements
generally agree to within these errors.  Given this agreement, it is
likely that results of the joint analysis of the \emph{HST} and TGAS
data, as reported in Tables~\ref{t:param} and~\ref{t:paramS}, yield
the most accurate estimates to date.

The values of the TGAS COM PMs are not strongly dependent on the
details of our PM field models. Since the stars are distributed
more-or-less symmetrically around the center, a mean of the PM data
(Table~\ref{t:TGAS}) yields results that are similar to our best fits
at the level of the random errors. The exception is the mean $\mu_W$
for the LMC, which is affected by the paucity of TGAS stars on the
southeast side.

We have not explicitly included possible {\it spatial} correlations in
TGAS PM errors (Gaia Collaboration \etal 2016b; Lindegren \etal 2016)
in our analysis. The effect of such correlations would be to
underestimate the random error in the weighted average PM of a stellar
sample (Kroupa \& Bastian 1997). The agreement between our TGAS
results and the \emph{HST} results implies that any residual
systematics introduced by this must be below the random errors.

We also considered Tycho-2 stars in the TGAS catalog. The Tycho-2
catalog goes fainter, to $V \sim 14$, than the \emph{Hipparcos}
catalog, but the typical TGAS PM uncertainty $\gtrsim 1 \masyr$ is
much worse. We selected Tycho-2 stars in the areas of the LMC and SMC,
with LOS velocities from the SIMBAD database or the RAVE survey
(Kordopatis \etal 2013) that are consistent with LMC or SMC
membership. Stars with discrepant PMs were excluded. This yielded 210
LMC and 34 SMC stars, which still includes possible remaining
foreground contamination. We found this insufficient to obtain an
accurate independent estimate of the LMC and SMC COM PMs.
 
Analysis of the LMC PM data well defines the LMC dynamical center
(Table~\ref{t:param}) and yields a result that is consistent with the
average dynamical center from HI measurements. For the SMC, the
location of the stellar dynamical center is not well known a priori,
and it is not well constrained by the PM data. We adopt the result for
the HI center as our best estimate, as in K13, but
Table~\ref{t:paramS} shows that uncertainty in this center introduces
$\sim 0.1 \masyr$ PM uncertainty. This is less pronounced when the
TGAS and \emph{HST} PM data are analyzed jointly.

The viewing angles $(i,\Theta)$ of the LMC disk are not accurately
known, with different methods yielding results that differ at the
level of tens of degrees. The new results in Table~\ref{t:param} are
within the range of what has been found by other studies (e.g., van
der Marel \& Cioni 2001; vdM02; van der Marel \etal 2009; vdMK14), but
are not sufficiently accurate to convincingly pin down the values of
these angles.

The LMC rotation curve inferred from the TGAS data is more useful than
that derived from the \emph{HST} PM data, because it pertains to a
single stellar population instead of a mixed population. The rotation
amplitude for $R \gtrsim 2 \kpc$ is consistent with that inferred from
LOS velocity measurements for young stars (Figure~\ref{f:rotcurve}),
further validating the accuracy of the TGAS data. However, at $R
\lesssim 2 \kpc$ the TGAS-inferred rotation curve lies somewhat below
the LOS rotation curve. This could be due to shot noise and the
limited number of (ten) TGAS stars at these radii, or it could reflect
shortcomings in our kinematical model for the young stellar disk (e.g.,
warping; Nikolaev \etal 2004).
  
The LMC PMs are measured in mas/yr, while LOS velocities are measured
in km/s. Comparison therefore yields a kinematic estimate of the
galaxy distance. The distance modulus implied by a joint fit is $m-M =
18.54 \pm 0.39$ (Table~\ref{t:param}, column~(5)). This is consistent
with the canonical $m-M = 18.50 \pm 0.10$ (Freedman \etal 2001), but
not competitive in terms of accuracy. This is due in part to the
random errors on the PM rotation curve, and in part to the random
errors on the inclination of the disk. But this method for determining
the LMC distance will become more competitive with future \emph{Gaia}
data releases.

The TGAS data for the SMC do not imply a significant rotation in its
young stellar population (see $V_{\rm rot}$ in Table~\ref{t:paramS}),
and certainly less than the $\sim 40 \kms$ rotation amplitude of HI
(Stanimirovic \etal 2004). Whether the young SMC stars show more
rotation than the old stars, as suggested by LOS velocity data (Evans
\& Howarth 2008; Harris \& Zaritsky 2006; Dobbie \etal 2014), remains
unclear.

\section{Conclusions}
\label{s:conc}

We have used the \emph{Gaia} DR1 to obtain new insights into the
motions and internal kinematics of the Magellanic Clouds. The results
do not improve upon the accuracy of existing \emph{HST} studies, but
they have similar accuracy and are consistent to within the
uncertainties. Since these missions use different methods with
different systematics, this provides an external validation of each
approach.\footnote{With future Gaia DRs it will be possible to do
  star-by-star comparisons, since $\gtrsim 100$ stars in the observed
  HST fields are bright enough for Gaia PM measurements (Kallivayalil
  et al.~2006a, fig.~6).} The TGAS results confirm the large PM of the
Magellanic Clouds, which has previously been used to revise our
understanding of their orbital history and cosmological context
(K13). Both \emph{Gaia} and \emph{HST} (vdMK14) confidently detect and
quantify the rotation of the LMC disk. Comparison of the LMC rotation
curves from PM and LOS data yields a kinematic distance estimate that
is independent from, but consistent with, that from photometric
methods and the cosmological distance ladder.

The results presented here are the first from the \emph{Gaia} mission
on the topic of Local Group dynamics. \emph{Gaia}'s future data releases will
contain many more stars and have higher PM accuracy. With the methods
used here, this is guaranteed to further improve our understanding of
the Magellanic Clouds. When combined with studies of other nearby
targets, this will revolutionize our understanding of the Milky Way
and its satellites. For PM studies further out into the Local Group, and
especially for dwarf galaxies with old stellar populations, \emph{HST}
will continue to be the telescope of choice due to its ability to
measure accurate PMs for faint stars ($V \lesssim 25$) over small
fields (e.g., Sohn \etal 2015).

\acknowledgements

R.v.d.M. is grateful to Nitya Kallivayalil and Gurtina Besla for
long-standing collaboration on the proper motion dynamics of the
Magellanic Clouds, and to all HSTPROMO members for discussions on
these and related topics.  J.S. was supported by an ESA Research
Fellowship in Space Science. This work has made use of data from the
ESA space mission \emph{Gaia} (\url{http://www.cosmos.esa.int/gaia}),
processed by the \emph{Gaia} Data Processing and Analysis Consortium
(DPAC,
\url{http://www.cosmos.esa.int/web/gaia/dpac/consortium}). Funding for
the DPAC has been provided by national institutions, in particular the
institutions participating in the \emph{Gaia} Multilateral Agreement.
Figure~1 was generated using ASTROPY (Robitaille \& Bressert 2012).

{\it Facilities:} \facility{\emph{Gaia}}.


{}

\end{document}